\documentclass[12pt]{article}

\usepackage{amsmath,verbatim,amssymb,epsfig,lscape}
\usepackage{graphicx,setspace,float,rotating,longtable}
\usepackage{color}

\def\singlespace{\def\baselinestretch{1}\@normalsize}

\def\boxit#1{\vbox{\hrule\hbox{\vrule\kern6pt
          \vbox{\kern6pt#1\kern6pt}\kern6pt\vrule}\hrule}}

\begin{document}
\baselineskip=24pt
\newcommand{\bbeta}{\boldsymbol{\beta}}
\newcommand{\btheta}{\boldsymbol{\theta}}
\newcommand{\bphi}{\boldsymbol{\phi}}
\newcommand{\bpsi}{\boldsymbol{\psi}}
\newcommand{\by}{\boldsymbol{y}}
\newcommand{\bY}{\boldsymbol{Y}}
\newcommand{\bm}{\boldsymbol{m}}
\newcommand{\bM}{\boldsymbol{M}}
\newcommand{\cD}{{\cal D}_{\rm obs}}
\newcommand{\blambda}{\boldsymbol{\lambda}}
\newcommand{\bs}{\boldsymbol}

\begin{center}
{\large \bf Bayesian Decision-optimal Interval Designs for Phase I Clinical Trials}
\end{center}

\begin{center}
Suyu Liu and Ying Yuan*

{\em
Department of Biostatistics\\
The University of Texas MD Anderson Cancer Center\\
Houston, Texas 77030, U.S.A. \\
*email: yyuan@mdanderson.org\\
}
\end{center}


\begin{center}
{\bf Abstract}
\end{center}

Interval designs are a class of phase I trial designs for which the decision of dose assignment is determined by comparing the  observed toxicity rate at the current dose with a prespecified (toxicity tolerance) interval. If the observed toxicity rate is located within the interval, we retain the current dose; if the observed toxicity rate is greater than the upper boundary of the interval, we deescalate the dose; and if the observed toxicity rate is smaller than the lower boundary of the interval, we escalate the dose. The most critical issue for the interval design is choosing an appropriate interval so that the design has good operating characteristics. By casting dose finding as a Bayesian decision-making problem, we propose new flexible methods to select the interval boundaries so as to minimize the probability of inappropriate dose assignment for patients. We show, both theoretically and numerically, that the resulting optimal interval designs not only have desirable finite- and large-sample properties, but also are particularly easy to implement in practice. Compared to existing designs, the proposed (local) optimal design has comparable average performance, but a lower risk of yielding a poorly performing clinical trial.  

\par\vspace{3mm}

%


\vspace{10mm}
\section{\textbf{Introduction}}
Numerous phase I trial designs have been proposed  to identify the maximum tolerated dose (MTD) of a new drug, which is typically defined as the dose with a dose-limiting toxicity probability that is closest to the target toxicity rate. The standard 3+3 design is easy to implement and widely used in practice, but suffers from poor operating characteristics (Storer, 1989; Korn et al., 1994; Ahn, 1998; and Iasonos et al., 2008). To obtain better operating characteristics, O'Quigley, Pepe and Fisher (1990) proposed the continual reassessment method (CRM), which utilizes the accrued information to continuously update the estimate of the parametric dose-toxicity model for dose assignment and selection. Durham and Flournoy (1994) proposed the biased coin design, which determines the dose assignment based on the toxicity outcome from the most recently treated patient. Whitehead and Brunier (1995) introduced a decision-theoretic approach geared to optimize information gathering for estimation of the MTD.  Wetherill (1963) proposed a $k$-in-a-row design that utilizes observations from the $k$ most recently treated patients to make the decision of dose assignment.  Babb, Rogatko and Zacks (1998) proposed a dose escalation method with overdose control that directly controls the probability of overdosing.  Lin and Shih (2001) investigated the statistical properties of the ``A+B" designs and derived the exact formulae for the corresponding statistical quantities.  Leung and Wang (2001) proposed an up-and-down design based on isotonic regression without making any parametric dose-response assumptions. Tsutakawa (1967) and Gezmu and Flournoy (2006) investigated the group up-and-down design for which the patients are treated in groups or cohorts. Chevret (2006), Ting (2006)  and Gerke and Siedentop (2008) provided comprehensive reviews of dose-finding methods for phase I clinical trials.

The interval design, a term coined by Oron, Azriel and Hoff (2011), is a relatively new class of phase I trial designs, for which the dose transition is defined by the relative location of the observed toxicity rate (i.e., the number of patients who experienced toxicity divided by the total number of patients treated) at the current dose with respect to a prespecified toxicity tolerance interval. If the observed toxicity rate is located within the interval, we retain the current dose; if the observed toxicity rate is greater than the upper boundary of the interval, we deescalate the dose; and if the observed toxicity rate is smaller than the lower boundary of the interval, we escalate the dose. Yuan and Chappell (2004) first suggested using a (tolerance) interval to determine dose escalation and deescalation.  Ivanova, Flournoy and Chung (2007) developed a more formal interval design, called the cumulative cohort design. That design was further extended to ordinal and continuous outcomes based on a $t$-statistic (Ivanova and Kim, 2009).  The modified toxicity probability interval design proposed by Ji et al.(2010) also utilized the idea of a tolerance interval to divide the posterior distribution of the toxicity probability of the current dose into three regions for dose finding.

One advantage of interval designs is that they are very simple to implement in practice. Because the interval is prespecified, during trial conduct, the decision of which dose to administer to the next cohort of patients does not require complicated computations, but only a simple comparison of the observed toxicity rate at the current dose with the prespecified interval boundaries. 
Compared to other similarly easy-to-implement designs, e.g., up-and-down designs, the interval design is more efficient because it uses all the information from the cumulative data at the current dose to determine the next dose assignment. Recently, Oron, Azriel and Hoff (2011) showed that the interval design provides a convergence property similar to that of the CRM. The interval design converges almost surely, at a $\sqrt{n}$ rate, to exclusive allocations at a dose level for which the true toxicity rate is within the interval.

Under the interval design, the selection of an appropriate tolerance interval is critical for the performance of the design because the dose transition and assignment are entirely determined by this prespecified interval.  Ivanova, Flournoy and Chung (2007) suggested selecting the interval by an exhaustive numerical search over the group up-and-down designs based on its large-sample property.
In this article, motivated by clinical practice, we propose a more flexible, finite-sample based approach to selecting the interval boundaries,
and use it to develop two optimal interval designs.

The idea behind our method is straightforward. The conduct of phase I trials can be essentially viewed as a sequence of decision-making steps of dose assignment for patients who are sequentially enrolled into the trial. At each moment of decision making, based on the observed data, we take one of three actions: escalate, deescalate or retain the current dose. Under the standard assumption that efficacy monotonically increases with toxicity,
an ideal trial design would escalate the dose when the current dose is below the MTD in order to avoid treating a patient at subtherapeutic dose levels; deescalate the dose when the current dose is above the MTD in order to avoid exposing a patient to overly toxic doses; and retain the same dose level when the current dose is equal (or close) to the MTD. 
However, such an ideal design is not available in practice because we do not know whether the current dose is actually below, above or equal (or close) to the MTD, and need to infer that information and make decisions based on the data collected from patients who have been enrolled in the trial. Given the randomness of the observed data and small sample sizes of phase I trials, the decisions of dose assignment we make are often incorrect, e.g., we may escalate (or deescalate) the dose when it  is actually higher (or lower) than the MTD, which results in overly aggressive (or conservative) dose escalation and treating excessive numbers of patients at dose levels above (or below) the MTD. From a practical and ethical viewpoint, it is highly desirable to minimize such decision errors so that the actual design behaves as closely as possible to the ideal (error-free) design.
Motivated by this goal, we cast dose finding as a sequential decision-making problem, and develop interval designs that minimize the expected decision error rate for trial conduct. The proposed designs are simpler to implement than the CRM, and the simulation study shows that they yield good operating characteristics that are comparable to or better than those of the CRM.

The remainder of this article is organized as follows. In Section 2,
we formulate dose finding as a sequential decision-making problem and propose designs to minimize the decision errors of dose assignment during trial conduct.  In Section 3, we present simulation studies to compare the operating characteristics of the new designs with those of some available designs. We conclude with a brief discussion in Section 4.

\section{Method}
\subsection{Interval design}
Assume that a total of $J$ prespecified doses are under investigation, and let $\phi$ denote the target toxicity rate specified by physicians. We assume that patients are treated in cohorts but allow the cohort size to vary from one cohort to another. Interval designs can be generally described as follows:
\begin{enumerate}
\item Patients in the first cohort are treated at the lowest dose level.
\item At the current dose level $j$, assume that a total (or the cumulative number) of $n_j$ patients have been treated, and $m_j$ of them have experienced toxicity. Let $\hat{p}_j=m_j/n_j$ denote the observed toxicity rate at dose level $j$, and $\lambda_{1j}(n_j, \phi)$ and $\lambda_{2j}(n_j, \phi)$  denote the prespecified lower and upper (or dose escalation and deescalation) boundaries of the interval, respectively, with $0 \le \lambda_{1j}(n_j, \phi)<\lambda_{2j}(n_j, \phi)\le 1$. To assign a dose to the next cohort of patients,
\begin{itemize}
\item if $\hat{p}_j \le \lambda_{1j}(n_j, \phi)$, we escalate the dose level to $j+1$;  
\item if $\hat{p}_j \ge \lambda_{2j}(n_j, \phi)$, we deescalate the dose level to $j-1$; 
\item otherwise, i.e., $ \lambda_{1j}(n_j, \phi) < \hat{p}_j < \lambda_{2j}(n_j, \phi)$, we retain the same dose level, $j$. 
\end{itemize}
To ensure that the dose levels of the treatment always remain within the prespecified dose range, the dose escalation/deescalation rule requires some adjustments when $j$ is at the lowest or highest level. That is, if $j=1$ and  $\hat{p}_j\ge \lambda_{2j}(n_j, \phi)$ or  $j=J$ and $\hat{p}_j \le \lambda_{1j}(n_j, \phi)$, the dose remains  at the same level, $j$. 

\item This process is continued until the maximum sample size is reached. 
\end{enumerate}
This design is more general than the existing interval designs (e.g., Ivanova, Flournoy and Chung, 2007; Oron, Azriel and Hoff, 2011), in the sense that it allows the interval boundaries $\lambda_{1j}(n_j, \phi)$ and $\lambda_{2j}(n_j, \phi)$  to depend on both dose level $j$ and $n_j$ (i.e., the number of patient that have been treated at dose level $j$), whereas the existing interval designs assume that the interval boundaries are independent of $j$ and $n_j$.

\subsection{Local optimal interval design} 

In the interval design, the selection of the interval boundaries $\lambda_{1j}(n_j, \phi)$ and $\lambda_{2j}(n_j, \phi)$ is critical because these two parameters essentially determine the operating characteristics of the design. To simplify the notations, we suppress the arguments $n_j$ and $\phi$ in  $\lambda_{1j}(n_j, \phi)$ and $\lambda_{2j}(n_j, \phi)$. In the following subsection, we describe a method to select $\lambda_{1j}$ and $\lambda_{2j}$ to minimize incorrect decisions of dose escalation and deescalation during the trial conduct. For convenience, we call the resulting design the local optimal interval design because the optimization is anchored to three point (or local) hypotheses.

\subsubsection{Minimizing the local decision error} 

In order to minimize incorrect decisions of dose assignment, we first formally define the correct and incorrect decisions.
Letting $p_j$ denote the true toxicity probability of dose level $j$ for $j=1, \ldots, J$, we formulate three point hypotheses:
\begin{eqnarray}
H_{0j}: &&  p_j = \phi  \notag \\
H_{1j}: && p_j = \phi _1 \label{pointh}\\
H_{2j}: && p_j = \phi_2, \notag
\end{eqnarray}
where $\phi _1$ denotes the highest toxicity probability that is deemed subtherapeutic (i.e., below the MTD) such that dose escalation should be made, and $\phi_2$ denotes the lowest toxicity probability that is deemed overly toxic such that dose deescalation is required. For simplicity, we will
alter the notation slightly by dropping subscript $j$ from $H_{0j}$, $H_{1j}$ and $H_{2j}$ wherever such alterations do not cause confusion.

Specifically, $H_{0}$ indicates that the current dose $d_j$ is the MTD and we should retain the current dose to treat the next cohort of patients; $H_{1}$ indicates that the current dose is subtherapeutic (or below the MTD) and we should escalate the dose; and $H_{2}$ indicates that the current dose is overly toxic (or above the MTD) and we need to deescalate the dose. Therefore, the correct decisions under $H_{0}$, $H_{1}$ and $H_{2}$ are retainment, escalation and deescalation (each based on the current dose level), denoted as ${\cal R}$, ${\cal E}$ and ${\cal D}$, respectively. Correspondingly, the incorrect decisions under $H_0$, $H_1$ and $H_2$ are $\bar{\cal R}$, $\bar{\cal E}$ and $\bar{\cal D}$, respectively, where $\bar{\cal R}$ denotes the decisions complementary to ${\cal R}$ (i.e., $\bar{\cal R}$ includes ${\cal E}$ and ${\cal D}$), and $\bar{\cal D}$ and $\bar{\cal R}$ denote the decisions complementary to ${\cal D}$ and  ${\cal R}$. 

We note that herein the purpose of specifying three hypotheses, $H_0, H_1$ and $H_2$, is not to represent the truth and conduct hypothesis testing, but just to indicate the cases of special interest under which we optimize the performance of our design. In particular, $H_1$ and $H_2$, or more precisely $\delta_1=\phi -\phi_1$ and $\delta_2=\phi_2-\phi$, represent the  minimal differences (or effect sizes) of practical interest to be distinguished from the target toxicity rate $\phi$ (or $H_0$), under which we want to minimize the average decision error rate for the trial conduct. A difference smaller than $\delta_1$ and $\delta_2$ may not be of practical importance, and a larger difference will lead to a smaller error rate because it is more easily distinguished from $\phi$ than  $\phi _1$ and $\phi_2$. This approach is analogous to sample size determination, for which we first specify a point alternative hypothesis to represent the minimal effect size of interest and then determine the sample size to ensure a desirable power under that hypothesis.

Under the Bayesian paradigm, we assign each of the hypotheses a prior probability of being true, denoted as $\pi_{kj} = {\rm pr}(H_{kj}), k=0, 1, 2$. Then under the proposed design, the probability of making an incorrect decision (the decision error rate), denoted as $\alpha(\lambda_{1j}, \lambda_{2j})$, at each of the dose assignments is given by
{
\begin{eqnarray}
&&\alpha(\lambda_{1j}, \lambda_{2j})\notag \\
 &=& {\rm pr}(H _{0j}) {\rm pr}(\bar{\cal R}| H _{0j}) +{\rm pr}(H _{1j}) {\rm pr}(\bar{\cal E} | H _{1j}) +{\rm pr}(H _{2j}) {\rm pr}(\bar{\cal D} | H _{2j}) \label{lerror} \\
&=& {\rm pr}(H _{0j}) {\rm pr}(m_j \le n_j\lambda_{1j} \, {\rm or} \, m_j\ge n_j\lambda_{2j}  | H _{0j}) + {\rm pr}(H _{1j}) {\rm pr}(m_j > n_j\lambda_{1j} | H _{1j}) \notag \\
&& +{\rm pr}(H _{2j}) {\rm pr}(m_j < n_j\lambda_{2j} | H _{2j}) \notag  \\
&=& \pi_{0j} \{ Bin(n_j\lambda_{1j}; n_j, \phi) + 1 - Bin(n_j\lambda_{2j}-1; n_j, \phi) \} + \pi_{1j} \{1-Bin(n_j\lambda_{1j}; n_j, \phi _1)\} \notag \\
&&+\pi_{2j} Bin(n_j\lambda_{2j}-1; n_j, \phi_2) \notag 
\end{eqnarray}
}
where $Bin(b; n, \phi)$ is the cumulative density function of the binomial distribution with size and probability parameters $n$ and $\phi$, evaluated at the value $b$.  It can be shown that the  decision error rate $\alpha(\lambda_{1j}, \lambda_{2j})$ is minimized when
\begin{eqnarray}
\lambda_{1j}  &=& \frac{\displaystyle {\rm log}\left(\frac{1-\phi _1}{1-\phi}\right)+ n_j ^{-1}{\rm log}\left(\frac{\pi_{1j}}{\pi_{0j}}\right)  }{\displaystyle{\rm log}\left(\frac{\phi(1-\phi _1)}{\phi_1(1-\phi)}\right)} \label{llambda1} \\
\lambda_{2j} &=& \frac{\displaystyle{\rm log}\left(\frac{1-\phi}{1-\phi_2}\right) + n_j^{-1}{\rm log}\left(\frac{\pi_{0j}}{\pi_{2j}}\right) }{\displaystyle{\rm log}\left(\frac{\phi_2(1-\phi)}{\phi(1-\phi_2)}\right)}.\label{llambda2}
\end{eqnarray}

The derivation of this result (see the Appendix) reveals that the values of $\lambda_{1j}$ and $\lambda_{2j}$ are actually the boundaries at which the posterior probabilities of $H_1$ and $H_2$, respectively, become more likely than that of $H_0$, i.e., $\lambda_{1j} = {\rm argmax}_{\hat{p}_j}({\rm pr}(H_1|n_j, m_j) > {\rm pr}(H_0|n_j,  m_j))$  and $\lambda_{2j} = {\rm argmin}_{\hat{p}_j}({\rm pr}(H_2|n_j, m_j) > {\rm pr}(H_0|n_j, m_j))$. This intuitive interpretation of $\lambda_{1j}$ and $\lambda_{2j}$ provides a natural justification for our dose assignment rule. That is, we should escalate the dose if $\hat{p}_j\le \lambda_{1j}$ because the observed data support that $H_1$ is more likely than $H_0$ (i.e., the current dose $j$ is below the MTD); and we should deescalate the dose if $\hat{p}_j \ge\lambda_{2j}$ because the observed data support that $H_2$ is more likely than $H_0$ (i.e., the current dose $j$ is above the MTD). 

\begin{figure}
\setlength{\abovecaptionskip}{-5pt}
\setlength{\belowcaptionskip}{5pt}
\begin{center}
\includegraphics[angle=0,width=5 in, totalheight=4 in] {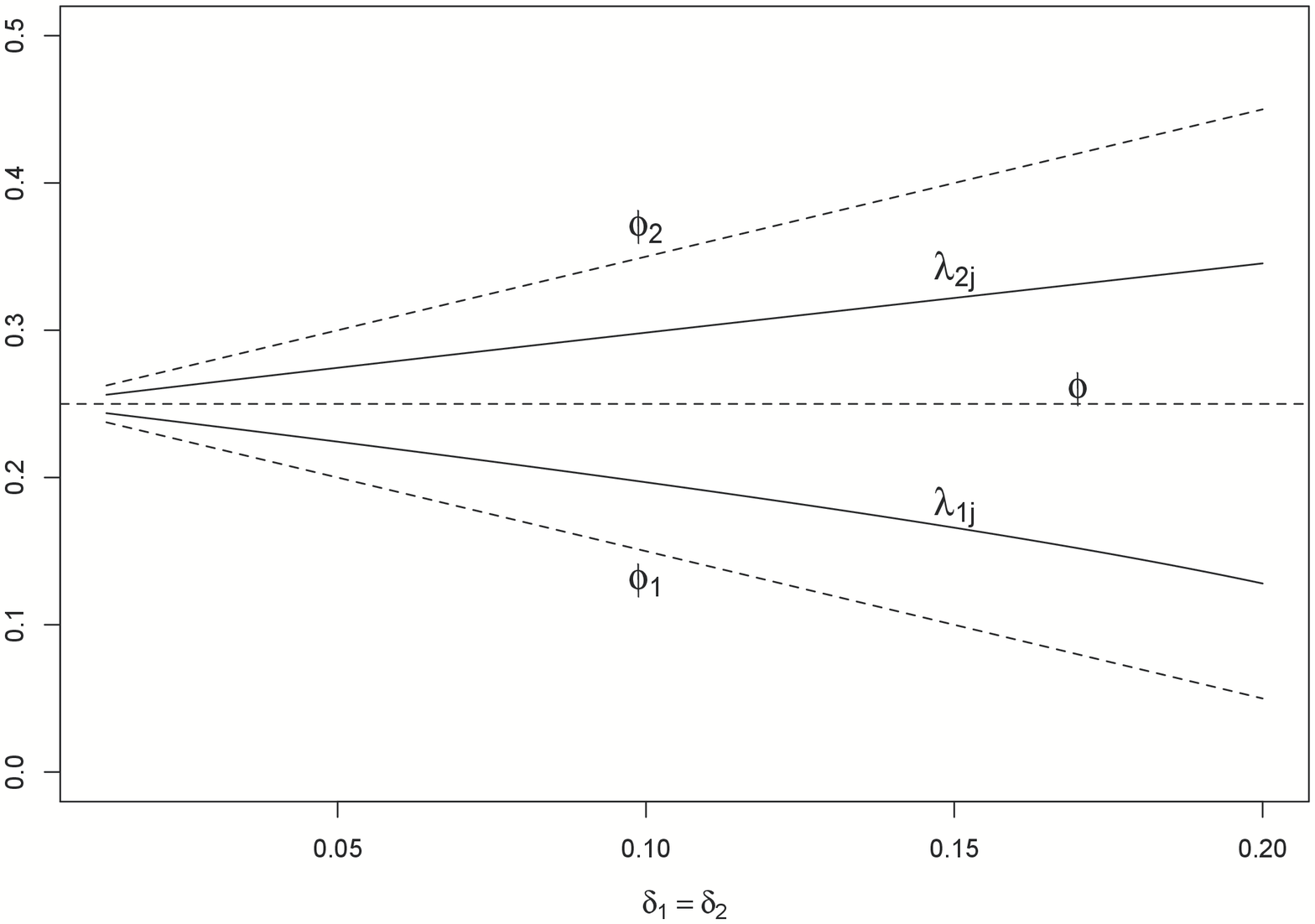}
\end{center}
\caption{Interval boundaries $\lambda_{1j}$ and $\lambda_{2j}$ changing with the values of $\phi_1$ and $\phi_2$ when the target toxicity rate $\phi=0.25$ and $\pi_{0j}=\pi_{1j}=\pi_{2j}$. \label{figlocal}}
\end{figure}

In addition, if we assign equal prior probabilities to three hypotheses (i.e., $\pi_{0j} = \pi_{1j} =\pi_{2j} =1/3$), the values of $\lambda_{1j}$ and $\lambda_{2j}$ simply become the likelihood-ratio hypothesis-testing boundaries. In this case, the value of $\lambda_{1j}$ is always located between $\phi_1$ and $\phi$ and the value of $\lambda_{2j}$ is always located between $\phi$ and $\phi_2$ (i.e., $\phi_1<\lambda_{1j}<\phi$ and $\phi<\lambda_{2j}<\phi_2$). Figure \ref{figlocal} shows how $\lambda_{1j}$ and $\lambda_{2j}$ change with respect to $\phi_1$ and $\phi_2$ when $\phi=0.25$. We can see that when $\phi_1$ and $\phi_2$ are symmetric around $\phi$ (i.e., $\delta_1=\delta_2$), the resulting $\lambda_{1j}$ and $\lambda_{2j}$ are also close to, although not exactly symmetric around $\phi$. Interestingly, based on equations (\ref{llambda1}) and (\ref{llambda2}), it is easy to see that when $\pi_{0j} = \pi_{1j} =\pi_{2j}$, $\lambda_{1j}$ and $\lambda_{2j}$ are invariant to both dose level $j$ and the accumulative sample size $n_j$. This property is appealing in practice because in this case the same pair of interval boundaries can be conveniently used throughout the trial, regardless of the dose level and the number of patients treated in the trial, which thus greatly simplifies the trial conduct.

\subsubsection{Design properties} 

Cheung (2005) introduced the concept of coherence for trial design and defined it as the design property by which dose escalation (or deescalation) is prohibited when the observed toxicity rate in { the most recently treated cohort} is more (or less) than the target toxicity rate.
Because that definition is based on the response from only the most recently treated cohort, ignoring the responses from cohorts previously enrolled and treated, herein we refer to it as {\it short-memory coherence}. As an extension, we define {\it long-memory coherence} as the design property by which dose escalation  (or deescalation) is prohibited  when the observed toxicity rate in { the accumulative cohorts} at the current dose is  more (or less) than the target toxicity rate. From a practical view point, long-memory coherence is a more relevant design property for conducting clinical trials than short-memory coherence. The greater relevance can be seen in the practical judgment of clinicians when they determine whether dose escalation (or deescalation) is plausible. It is preferable for clinicians to base their decision on the toxicity data from all patients treated at the current dose rather than from just 2 or 3 patients in the most recently treated cohort, which could provide highly variable observations due to the small sample size of the cohort (i.e., rarely more than 3) and patient heterogeneity. As shown in the Appendix, the proposed local optimal interval design has the following finite-sample property.\\

{\noindent\bf Theorem 1.} When setting $\pi_{0j} = \pi_{1j} =\pi_{2j}$, the proposed local optimal interval design is long-memory coherent in the sense that the probability of dose escalation (or deescalation) is zero when the observed toxicity rate  $\hat{p}_j$ at the current dose  is higher (or lower) than the target toxicity rate $\phi$. \\

This finite-sample property makes the local optimal design very appealing in practice because it automatically satisfies the following (ad hoc) safety requirement often imposed by clinicians: dose escalation is not allowed if the observed toxicity rate at the current dose is higher than the MTD.  
Because the local optimal interval design uses the toxicity information from  previously treated patients to make the decision of dose transition, it is not short-memory coherent. We note that the commonly used version of the CRM, which does not allow a dose to be skipped, is not short-memory coherent either, especially at the beginning of the trial and when using an informative prior.

We now turn to another important property of the design --- the MTD selection. 
Under the proposed optimal dose assignment, we tend to treat patients at (or close to) the MTD, which facilitates the MTD selection at the end of the trial because most data and statistical power are concentrated around the MTD.  
Since $\lim_{n_j \to \infty} \lambda_{1j} = {\rm log}(\frac{1-\phi_1}{1-\phi})/{\rm log}(\frac{\phi(1-\phi_1)}{\phi_1(1-\phi)}) \equiv \tilde{\lambda}_{1}$ and $\lim_{n_j \to \infty} \lambda_{2j} = {\rm log}(\frac{1-\phi}{1-\phi_2})/{\rm log}(\frac{\phi_2(1-\phi)}{\phi(1-\phi_2)})\equiv \tilde{\lambda}_{2}$, it can be shown that the proposed design has the following asymptotic dose-selection property. \\

{\noindent\bf Theorem 2.} Dose allocation in the local optimal interval design converges almost surely to dose level ${j^*}$ if $p_{j^*} \in (\tilde{\lambda}_1, \tilde{\lambda}_2)$ and dose level $j^*$ is the only dose satisfying $p_{j^*} \in [\tilde{\lambda}_1, \tilde{\lambda}_2]$. If no dose level satisfies $p_{j} \in (\tilde{\lambda}_1, \tilde{\lambda}_2)$ but $\phi \in [p_1, p_J]$, the local optimal interval design would eventually oscillate almost surely between the two dose levels at which the associated toxicity probabilities straddle the target interval. If there are multiple dose levels satisfying $p_{j} \in (\tilde{\lambda}_1, \tilde{\lambda}_2)$, the local optimal interval design will converge almost surely to one of these levels. \\

The proof of this result is similar to that of Oron, Azriel and Hoff (2011), noting that $\lambda_{1j}$ and $\lambda_{2j}$ converge to constants $\tilde{\lambda}_1$ and $\tilde{\lambda}_2$.

\subsubsection{Practical implementation} 
To use the local optimal interval designs in practice, we need to specify the values of $\phi _1$ and $\phi_2$.
In general, we should avoid setting the values of  
$\phi _1$ and $\phi_2$ very close to $\phi$. This is because the small sample sizes of typical phase I trials prevent us from differentiating the target toxicity rate from the rates close to it. In addition, in most clinical applications, the target toxicity rate is often a rough guess, and finding a dose level with a toxicity rate reasonably close to the target rate will still prove to be of interest to the investigator. Based on our experience with phase I oncology trials and the operating characteristics of the proposed design, we find that $\phi _1 \in [0.5\phi, 0.7\phi]$ and $\phi_2 \in [1.3\phi, 1.5\phi]$ are reasonable values for most clinical applications. As default values, we recommend $\phi _1 = 0.6\phi $ and $\phi_2 = 1.4\phi$ (i.e.,  40\% deviation from the target) for general use. For example, if the target toxicity is 0.25, we may set $\phi _1=0.15$ and $\phi_2=0.35$.

The other parameters we need to specify for the trial design are $\pi_{kj}$, the prior probability of the hypothesis $H_{kj}$, $k=0, 1, 2$. 
As $\pi_{kj}$'s represent the prior probabilities that dose level $j$ is below, equal to or above the MTD, their values can be directly elicited from physicians by asking them how likely it is that each of $J$ doses is below, equal to or above the MTD. When such prior information is not available, as is often the case in practice, we can take a ``noninformative" approach by assuming that the current dose has equal prior probabilities of being below, equal to or above the MTD. Since the interval design examines the dose levels one at a time (i.e., the current dose level) without borrowing information from other dose levels, and the current dose level can be at any level from 1 to $J$, this noninformative approach effectively results in the noninformative prior $\pi_{0j}=\pi_{1j}=\pi_{2j}=1/3$. 
One advantage of using such a noninformative prior is that the resulting interval boundaries $\lambda_{1j}$ and $\lambda_{2j}$ are invariant to dose level $j$ and the number of patients $n_j$.  Therefore, the same pair of boundaries can be conveniently used throughout the trial for dose assignment, regardless of the dose level and the number of enrolled patients.  Table \ref{tablocal} provides the interval boundaries for some commonly encountered target toxicity rates in oncology under the noninformative prior 
$\pi_{0j}=\pi_{1j}=\pi_{2j}=1/3$. It seems appealing to specify increasing prior probabilities for $\pi_{2j}$ (i.e., $\pi_{21}<\pi_{22}<\ldots, <\pi_{2J}$) to reflect that toxicity monotonically increases with dose levels. However, based on numerical studies, we found that such an approach did not improve the performance of the design (results are not shown). This may be because the monotonicity of the dose-toxicity relationship has been (implicitly) incorporated into the design through the dose escalation rule (i.e., we escalate the dose because we assume that the next higher dose level is more toxic), and thus using ordered priors will not bring enough extra information to improve the design performance. 

The local optimal interval design is very easy to implement in practice. Once we specify the design parameters as described above, the interval boundaries $\lambda_{1j}$ and $\lambda_{2j}$ can be easily calculated at the trial design phase based on equations (\ref{llambda1}) and (\ref{llambda2}).  Then during the trial conduct, clinicians can simply count the number of patients who experience toxicity and compare the observed toxicity rate $\hat{p}_j$ with the prespecified interval boundaries $\lambda_{1j}$ and $\lambda_{2j}$ to determine dose assignment until the trial is completed.

After the trial is completed, we need to select a dose level as the MTD. We propose to select the MTD based on $\{\tilde{p}_{j}\}$,  the isotonically transformed values of the observed toxicity rates $\{\hat{p}_{j}\}$. Specifically, we select as the MTD dose $j^*$, for which the isotonic estimate of toxicity rate $\tilde{p}_{j^*}$ is closest to $\phi$; if there are ties for $\tilde{p}_{j^*}$, we select from the ties the highest dose level when $\tilde{p}_{j^*}<\phi$ or the lowest dose level when $\tilde{p}_{j^*}>\phi$. The isotonic estimates $\{\tilde{p}_{j}\}$ can be obtained by applying the pooled--adjacent-violators algorithm (PAVA) (Barlow; 1972) to $\{\hat{p}_j\}$. Operatively, the PAVA replaces any adjacent $\hat{p}_j$'s that violate the nondecreasing order by their  (weighted) average so that the resulting estimates $\tilde{p}_j$'s become monotonic. In the case in which the observed toxicity rates are monotonic, $\tilde{p}_{j}$ and $\hat{p}_{j}$ are equivalent. We note that besides the above method, any other reasonable dose selection procedure can be used to select the MTD for the local optimal design, as the MTD selection and dose transition rules are two relatively independent components of the trial design, both conceptually and operatively.

One practical issue largely ignored in the previously proposed interval designs involves the risk of assigning too many patients to an overly toxic dose.  Because interval designs do not ``look" ahead, but instead use only the toxicity information at the current dose level to determine the dose escalation decision, the dose assignment will bounce back and forth between two adjacent doses when one of them is much lower than the MTD and the other is much higher than the MTD. To avoid this result of assigning too many patients to the overly toxic dose, we propose to impose the following dose elimination rule when implementing the local optimal interval design.
\begin{eqnarray*}
 &&\textrm{ If pr}(p_j>\phi|m_j, n_j)>0.95 \,\, \textrm{and}\,\, n_j \ge 3, \textrm{ dose levels {\it j} and higher are eliminated }  \label{stopping} \\
 && \quad \textrm{from the trial, and the trial is terminated if the first dose level is eliminated}, \notag
\end{eqnarray*}
where $\textrm{pr}(p_j>\phi|m_j, n_j)>0.95$ can be evaluated based on a beta-binomial model,  assuming that $m$ follows a binomial distribution (with size and probability parameters $n$ and $p_j$) and $p_j$ follows a vague beta prior, e.g., $p_j \sim beta(0.1, 0.1)$. Based on our experience, rather than repeatedly evaluating the above dose elimination rule in real time during the trial conduct, medical researchers often prefer to enumerate the dose elimination boundaries for each possible value of $n_j$ before the initiation of the trial and include these boundaries in the trial protocol. Therefore, when conducting the trial, they can determine dose elimination by simply examining whether the number of patients experiencing toxicity at the current dose, i.e., $m_j$, exceeds the elimination boundaries. Table 2 (in the bottom row) provides the elimination boundaries for $\phi=0.25$. For example, when the number of patients treated at the current dose $n_j=4$, we need to eliminate that dose and higher doses if 3 or more patients experience toxicity.

To facilitate practitioners applying the proposed designs, we have prepared easy-to-use software (written in R) to calculate the interval and dose elimination boundaries and select the MTD. The software is available upon request.

\subsection{Global optimal interval design} 

We have discussed the local optimal interval design, which minimizes the decision error under the three point hypotheses given by equation (\ref{pointh}). In this section, we discuss a global optimal interval design, which accounts for all possible values of $p_j$ by specifying three composite hypotheses. 
In the global optimal design, values of $\lambda_{1j}$ and $\lambda_{2j}$ are chosen to minimize the average decision error over the whole support of $p_j$, i.e., $p_j \in [0, 1]$. The contrast between the local and global optimal interval designs is somewhat analogous to the uniformly most powerful tests for simple hypotheses versus composite hypotheses in the frequentist testing framework. Specifically, we define three composite hypotheses, 
\begin{eqnarray*}
H_{0j}: && \phi _1 <p_j<\phi_2 \\
H_{1j}: && 0 \le p_j \le \phi _1 \\
H_{2j}: && \phi_2 \le p_j \le 1,
\end{eqnarray*}
where $H_{1j}$ indicates that dose level $j$ is subtherapeutic and we should escalate the dose; $H_{2j}$ indicates that dose level $j$ is overly toxic and we should deescalate the dose; and $H_{0j}$ indicates that dose level $j$ is close to the MTD and we should retain the same dose level. 

Under each hypothesis, we assign the toxicity probabilities $p_j$ a noninformative uniform prior: 
\begin{eqnarray}
f(p_j|H_{0j}) &=& {\rm Unif}(\phi _1, \phi_2) \nonumber \\
f(p_j|H_{1j}) &=& {\rm Unif}(0, \phi _1) \label{priorU}  \\
f(p_j|H_{2j}) &=& {\rm Unif}(\phi_2, 1). \nonumber
\end{eqnarray}
Then  the global decision error rate is given by
\begin{eqnarray}
&&{ \alpha_g(\lambda_{1j},\lambda_{2j})} \notag \\
 &=& {\rm pr}(H_{0j}) \int f(p_j|H_{0j}) {\rm pr}(\bar{\cal R}| p_j, H_{0j}) \,{\rm d} p_j+{\rm pr}(H_{1j}) \int f(p_j|H_{1j}) {\rm pr}(\bar{\cal E} | p_j, H_{1j})\,{\rm d} p_j \notag \\
 &&+{\rm pr}(H_{2j}) \int f(p_j|H_{2j})  {\rm pr}(\bar{\cal D} |p_j,  H_{2j})\,{\rm d} p_j  \label{gerror} \\
&=&  \pi_{0j} + \pi_{1j} + \sum_{y=0}^{b_{1j}}\left[ \frac{\pi_{0j}\{Beta(\phi_2; y+1, n_j-y+1)-Beta(\phi _1;y+1, n_j-y+1)\}}{(\phi_2-\phi _1)(n_j+1)} \right. \notag \\
&& \left. - \frac{\pi_{1j}Beta(\phi _1;y+1, n_j-y+1)}{\phi _1(n_j+1)} \right] \notag\\
&&  + \sum_{y=0}^{b_{2j}-1}\left\{\frac{\pi_{2j}Beta(\phi _2;y+1, n_j-y+1)}{(1-\phi_2)(n_j+1)}\right. \notag \\
&& \left. -  \frac{\pi_{0j}\{Beta(\phi_2;y+1, n_j-y+1)-Beta(\phi _1;y+1, n_j-y+1)\}}{(\phi_2-\phi _1)(n_j+1)}  \right\} \notag,
\end{eqnarray}
where $b_{1j}= {\rm floor}(n_j\lambda_{1j})$, $b_{2j}= {\rm floor}(n_j\lambda_{2j})$ and $Beta(c; a, b)$ is the cumulative distribution function of a beta distribution with the shape and scale parameters $a$ and $b$, evaluated at the value $c$. Although without closed forms, the following theorem can be shown to hold.  \\

{\noindent\bf Theorem 3.} The values of $\lambda_{1j}$ and $\lambda_{2j}$ that minimize the global decision error rate (\ref{gerror}) are  the boundaries at which the posterior probabilities of $H_1$ and $H_2$, respectively, become more likely than that of the $H_0$. \\

\noindent The proof of this result appears in the Appendix. In practice,  the values of $\lambda_{1j}$ and $\lambda_{2j}$ that minimize the global decision error rate can be easily determined by a numerical search, as $b_{1j}$ and $b_{2j}$ are integers between 0 and $n_j$.  

\begin{figure}
\setlength{\abovecaptionskip}{-5pt}
\setlength{\belowcaptionskip}{5pt}
\begin{center}
\includegraphics[angle=0,width=5 in, totalheight=4 in] {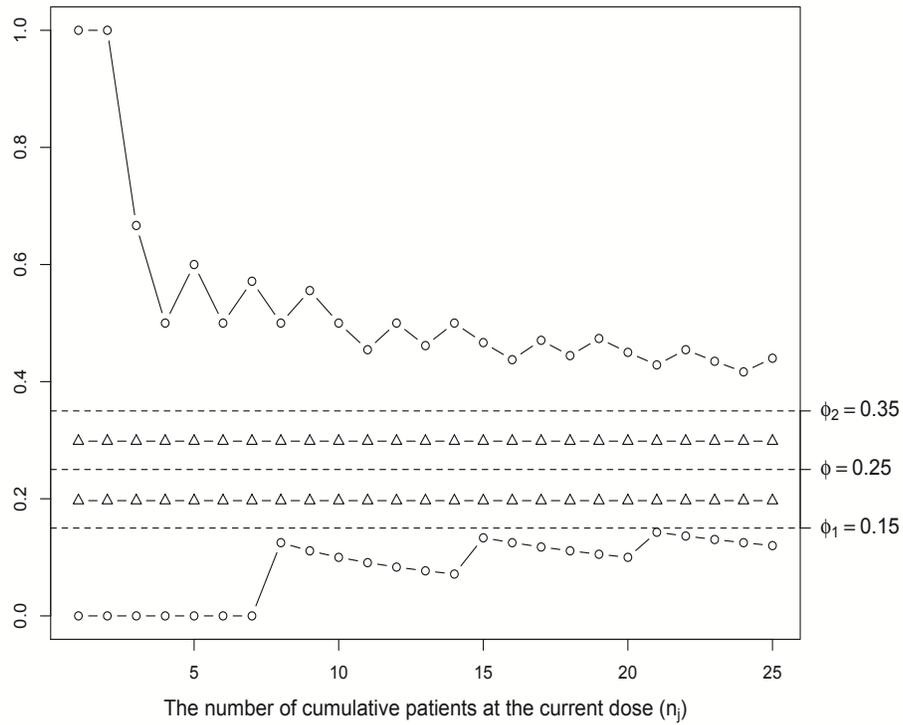}
\end{center}
\caption{Interval boundaries $\lambda_{1j}$ and $\lambda_{2j}$ for the local optimal design (triangles) and global optimal design (circles) under different numbers of cumulative patients at the current dose ($n_j$) when $\phi=0.25$, $\phi_1=0.15$, $\phi_2=0.35$ and $\pi_{0j}=\pi_{1j}=\pi_{2j}$. Under each design, the top line is $\lambda_{1j}$ and the bottom line is $\lambda_{2j}$. \label{figglobal}}
\end{figure}

Compared to the local optimal design,  the interval length (i.e., $\lambda_{2j}-\lambda_{1j}$) under the global optimal design is wider (see Figure \ref{figglobal}). This is because when we use the point-hypothesis values of the local optimal approach (i.e., $\phi_1$ and $\phi_2$) as the endpoints
of the composite hypothesis for the global approach, the latter is more favorable to the null because the most formidable competing hypotheses are then ``diluted" by a continuum that is also far less likely to contain ones. 
In addition, as shown in Figure \ref{figglobal},  unlike the local optimal design, for which $\lambda_{1j}$ and $\lambda_{2j}$ are invariant to $n_j$, for the global optimal design, $\lambda_{1j}$ and $\lambda_{2j}$ depend on $n_j$ even when the three hypotheses have equal prior probabilities. Despite these differences, we note that the global optimal design is also long-memory coherent because $\lambda_{1j}<\phi$ and $\lambda_{2j}>\phi$.\\

{\noindent\bf Theorem 4.} When setting $\pi_{0j} = \pi_{1j} =\pi_{2j}$, the global optimal design is long-memory coherent in the sense that the probability of dose escalation (or deescalation) is zero when the observed toxicity rate  $\hat{p}_j$ at the current dose  is higher (or lower) than the target toxicity rate $\phi$. \\

In addition, as shown in the Appendix, the global optimal design has the  convergence property described in Theorem 5.\\

{\noindent\bf Theorem 5}  Dose allocation in the global optimal interval design converges almost surely to dose level ${j^*}$ if $p_{j^*} \in (\phi_1, \phi_2)$ and dose level $j^*$ is the only dose satisfying $p_{j^*} \in [\phi_1, \phi_2]$. If no dose level satisfies $p_{j} \in (\phi_1, \phi_2)$ but $\phi \in [p_1, p_J]$, the global optimal interval design would eventually oscillate almost surely between the two dose levels at which the associated toxicity probabilities straddle the target interval. If there are multiple dose levels satisfying $p_{j} \in (\phi_1, \phi_2)$, the global optimal interval design will converge almost surely to one of these levels. \\

As for the practical implementation of this design, the same principle described in the previous section can be used to specify the design parameters $\phi_1$, $\phi_2$ and $\pi_{kj}$ for the global optimal design. The interval boundaries for the global optimal design do not have a closed form, but can be easily determined using a numerical search. The interval boundaries of the global optimal design depend on  $n_j$ even when we set $\pi_{0j}=\pi_{1j}=\pi_{2j}$, which makes the implementation of the design slightly more complicated than that of the local optimal design. That is, we need to find $\lambda_{1j}$ and $\lambda_{2j}$ for each possible value of $n_j$ before the initiation of the trial. Once these boundaries are determined, the global optimal design can be easily conducted by comparing the observed toxicity rate with the interval boundaries. Table \ref{tabglobal} displays the interval boundaries of the global optimal design when the target toxicity rate is 0.25 up to $n_j=15$. Thus for practical reasons, we may prefer to use the local optimal design. The finite sample simulation described in the next section also suggests that the local optimal design has  better operating characteristics.

\section{Simulation studies}
\begin{figure}
\setlength{\abovecaptionskip}{-5pt}
\setlength{\belowcaptionskip}{5pt}
\begin{center}
\includegraphics[angle=0,width=5 in, totalheight=4 in] {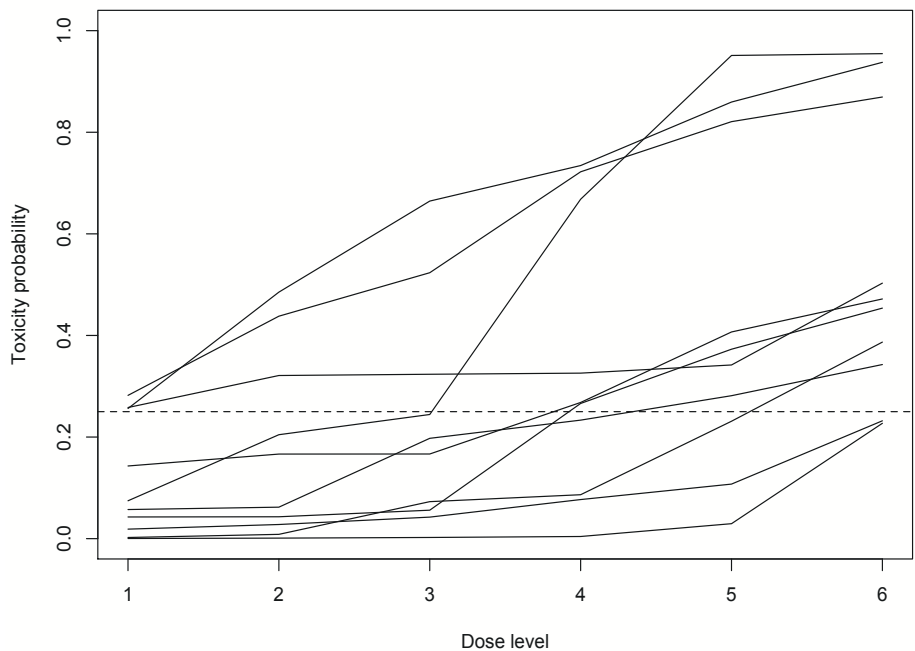}
\end{center}
\caption{Ten randomly generated dose-toxicity curves based on the simulation procedure described in Section 3. The dashed line indicates the target toxicity probability of 0.25. \label{dosetox}}
\end{figure}

We used simulation studies to compare the operating characteristics of the proposed designs with those of four available designs: the group up-and-down (GUD) design, the cumulative cohort design (CCD), the modified toxicity probability interval (mTPI) design, and the CRM.  
We assumed a total of $J=6$ dose levels with a sample size of $n=36$ patients in 12 cohorts of size 3. The target toxicity rate was $\phi=0.25$. In the proposed designs, we set $\phi_1=0.15$, $\phi_2=0.35$ and assigned equal prior probabilities to three hypotheses with $\pi_{0j}=\pi_{1j}=\pi_{2j}=1/3$. The same values of $\phi_1$ and $\phi_2$ were also used for the mTPI design. For the CCD, following Ivanova et al. (2007), we set the tolerance interval as $\phi \pm 0.09$. For the GUD design, we escalated the dose when no toxicities were observed in the 3-patient cohort, and deescalated the dose if any toxicities were observed in the 3-patient cohort, as recommended by Gezmu and Flournoy (2006) (i.e., the (3, 0, 1) design according to their terminology). The CRM was based on the model $p_j = a_j^{{\rm exp}(\alpha)}$ where the ``skeleton" $a_j = (0.01, 0.08, 0.25, 0.46, 0.65, 0.79)$ was chosen based on the model calibration method of Lee and Cheung (2009). We assigned the unknown parameter $\alpha$ the least-informative prior $\alpha \sim N(0, 1.24^2)$ as proposed by Lee and Cheung (2011), under which each dose level has an (approximately) equal prior probability of being the MTD.  Skipping a dose level was not allowed in the CRM. To make the designs more comparable, we applied the dose elimination rule and isotonic dose selection rule, as described in Section 2.2.3, to all designs except the CRM because that design has its own model-based safety and dose selection rules. We simulated 10,000 trials. To avoid cherry-picking scenarios that are better suited to specific methods, for each simulated trial, the toxicity scenario (i.e., the true toxicity probabilities of the six doses) was randomly generated based on the approach of Paoletti, O'Quigley and Maccario (2004), as follows: 
\begin{enumerate}
\item Randomly select, with equal probabilities, one of the $J$ dose levels as the MTD and denote this selected dose level by $j$. 
\item Generate  the toxicity probability of the MTD, $p_j=\Phi(\epsilon_j)$, where the random error $\epsilon_j \sim N(z(\phi), \sigma_0^2)$ with $\Phi(\cdot)$ and $z(\cdot)$ denoting the cumulative density function (CDF) and the inverse CDF of the standard normal distribution, respectively.
\item  Generate $p_{j-1}$ and $p_{j+1}$ (i.e., the toxicity probabilities of two doses adjacent to the MTD) under the constraint that $p_j$ is closest to $\phi$. This can be done by generating $p_{j-1} = \Phi [ z(p_j) - I_{\{z(p_j)>z({\phi})\}}\{z(p_j)-z(2\phi-p_j)\} - \epsilon_{j-1}^2]$ and $p_{j+1} = \Phi [ z(p_j) + I_{\{z(p_j)<z(\phi)\}}\{ z(2\phi-p_j)-z(p_j)\} + \epsilon_{j+1}^2]$ where $\epsilon_{j-1} \sim N(\mu_{1}, \sigma_1^2)$, $\epsilon_{j+1} \sim N(\mu_{2}, \sigma_2^2)$, and $I(.)$ is an indicator function. For example, $I_{\{z(p_j)>z({\phi})\}}$ equals 1 if $z(p_j)>z({\phi})$ and 0 otherwise.
\item Successively generate the toxicity probabilities for the remaining levels according to $p_{j-2} = \Phi [ z(p_{j-1}) - \epsilon_{j-2}^2]$ and $p_{j+2} = \Phi [ z(p_{j+1}) + \epsilon_{j+2}^2]$, and  $p_{j-3} = \Phi [ z(p_{j-2}) - \epsilon_{j-3}^2]$ and $p_{j+3} = \Phi [ z(p_{j+2}) + \epsilon_{j+3}^2]$, and so on, where $\epsilon_{j-2},  \epsilon_{j-3} \sim N(\mu_{1}, \sigma_1^2)$ and $\epsilon_{j+2}, \epsilon_{j+3} \sim N(\mu_{2}, \sigma_2^2)$.
\end{enumerate}
In our simulation, we set $\sigma_0=0.05$, $\sigma_1=\sigma_2=0.35$, and controlled the average probability difference around the target to be 0.07, 0.1 and 0.15 by varying the values of $\mu_1$ and $\mu_2$. For illustration, Figure \ref{dosetox} shows 10 (out of 10,000) randomly generated dose-toxicity curves when the average probability difference around the target was 0.1. We can see that these curves demonstrate various shapes and locations of the MTD.

Table \ref{OC} shows the simulation results, including the selection percentage of the MTD, the average percentage of patients treated at the MTD, the average toxicity rate, and the average sample size. In addition, we reported two risk measures for the designs: (1) the risk of poor allocation, defined as the percentage of simulation runs in which the number of patients allocated to the MTD (say $n_{_{MTD}}$) is less than that of a standard non-sequential design (i.e., $n_{_{MTD}}<n/J$); and (2) the risk of high toxicity, which is defined as the percentage of simulation runs in which the total number of toxicities is greater than $n\phi$. These risk measures are of great practical importance because they gauge the likelihood of a trial turning out to be a ``bad" trial (e.g., performing worse than a standard non-sequential trial) under a specific design.

The simulation results show that all the designs, except the GUD design, generally have a similar ``average" level of performance regarding several factors, including the MTD selection percentage, the average percentage of patients treated at the MTD, and the average toxicity rate. Among the six designs under comparison, the CCD,
mTPI, and global optimal designs have particularly similar performance levels. One reason for such similarity is that the decisions of dose transition in these designs are all based on the data observed at the current dose. Given that the number of patients treated at a specific dose level is typically very small (e.g., up to a few cohorts) and the toxicity outcome is discrete, the possible outcome patterns are actually very limited. Consequently, although based on different statistical criteria, these designs often end up with similar dose transition decisions and average performance levels, especially after averaging across a large number of random scenarios. The GUD design performs differently, yielding MTD selection percentages and average toxicity rates comparable to those of the other designs, but allocating lower percentages of patients to the MTD. For example, the percentages of patients allocated to the MTD under the GUD design are about 5\% and 10\% lower than those of the other designs when the respective average probability differences around the target are 0.1 and 0.15. 
This is because the GUD design never remains at the same dose level, except when possibly allocating patients to the lowest or highest dose. Therefore, it has a lower chance of staying at the MTD even when the observed data indicate that the current dose is the MTD. 

However, in terms of the risk of poor allocation decisions, the local optimal design and GUD design stand out and substantially outperform the other competing designs. For example, under the local optimal design, the risk of poor allocation is 14.3\% and 16.1\% lower than that under the CRM, and is 10.4\% and 10.8\% lower than that under the CCD when the average probability difference around the target is 0.1 and 0.07. For the risk of high toxicity, the proposed local optimal design and GUD design also perform better than the other designs. These results suggest that the use of the local optimal interval design decreases the likelihood of conducting a poorly performing trial compared to the use of the other designs.  Such an improvement is of great practical importance because we rarely run a trial more than a few times. What really concerns us is the likelihood of a clinical trial allocating too few patients to the MTD or resulting in too many toxicities, not the trial design's average performance over thousands of runs,  such as in a simulation study. Although the GUD design has the lowest risk of poor allocation decisions, unfortunately on average it allocates lower percentages of patients to the MTD than the other designs, as noted previously.
  
The CCD, global optimal design and local optimal design share similar interval-based dose transition schemes, so differences in the design risks (of poor allocation and high toxicity) should be due to the different choices of the interval width (i.e., $\lambda_{1j}$ and $\lambda_{2j}$). The interval width of the CCD (i.e., $\pm$0.9) is wider than that of the local optimal design (i.e., about 0.5), and is chosen based on a numerical search to ensure the convergence of the design. The wide (tolerance) interval of the CCD may restrict the freedom of dose movement and consequently lead to a higher risk of staying at doses that are not close to the MTD. This may also explain why the global optimal design has a higher risk of poor allocation decisions and high toxicity than the local optimal design, as the former also has a wider interval than the local optimal design (see Figure 2) due to the optimization over the whole support of $p_j$. The CRM is a model-based design that determines dose assignment based on the estimate of a parametric dose-toxicity model. Oron et al. (2011) and Azriel (2012) reported that under randomly-simulated dose-toxicity scenarios, the CRM often failed to provide coverage of the MTD (up to 90\% of the time), which can induce extra variations in the performance of the CRM (e.g., a higher risk of poor allocation decisions). 

The above random-scenario-based simulation provides an objective comparison of different designs by averaging over a large number of plausible scenarios. However, it cannot inform us of the performance of the designs under a specific toxicity scenario. Therefore, we conducted another simulation study based on a set of prespecified dose-toxicity scenarios with different locations of the MTD.  The simulation results are shown in Table \ref{fixedsimu}. Albeit there is some variation, the general pattern is similar to that under the random selection of simulated toxicity scenarios. That is, different designs yield roughly comparable ``average" performance levels (i.e., the MTD selection percentage and the average number of patients allocated to the MTD), but the proposed local optimal design and the GUD design have lower risks of poor allocation decisions or high toxicity than the other designs. Compared to the local optimal design, the drawback of the GUD design is that on average it allocates fewer patients to the MTD. One phenomenon that is more explicitly revealed by the fixed-scenario simulation is that the relative performance of the CRM depends on how much the assumed model structure (e.g., skeleton) deviates from the truth. The CRM performs better than the other designs if the target dose is close to its prior estimate of the MTD (i.e., dose level 3, for example, under scenarios 2 and 4), but performs worse than the other designs if the target dose deviates from its prior estimate (e.g., under scenarios 1 and 5).

\section{Conclusion}
We have proposed a flexible framework for constructing interval designs by treating phase I dose finding as a decision-making problem. We proposed local and global optimal interval designs that minimize the decision error of dose assignment based on either point or composite hypotheses.  We showed that the proposed designs have sound theoretical properties and good numerical performance. The proposed local and global optimal designs are substantially easier to implement than the CRM, yet they yield performances that are comparable to that of the CRM. 

We have considered the local and global optimal  designs, but the proposed decision-making framework is very flexible. These designs can be easily modified to accommodate different design objectives by specifying an appropriate object function to be minimized. For example, noting that the local and global decision errors given in equations (\ref{lerror}) and (\ref{gerror}) actually consist of different types of decision errors, we can propose minimax designs to prevent the rate of any individual type of decision error from being too high. 

In some cases, for safety reasons, we may be more concerned with incorrect dose escalation than with  incorrect dose deescalation. In these cases, we can classify the decision error, such as equation (\ref{lerror}), into errors of making incorrect  decisions of escalation, deescalation and dose level retainment. We then assign the appropriate weight to each type of error to reflect its relative importance, and minimize the resulting objective function. 

The proposed optimal interval designs use only the information observed from patients treated at the current dose to make the decision of dose assignment, resulting in a particularly easy-to-implement design structure. Somewhat surprisingly, our simulation studies show that ignoring data information at other doses does not have a noticeable impact on the performance of the design. The proposed designs on average perform as well as the CRM, which utilizes all the data observed across the dose range based on a parametric dose-response model. This phenomenon may have two explanations.  On one hand, although the optimal interval design uses only local information for dose assignment, it does use all the observed data (across doses) to select the MTD at the end of the trial. On the other hand, using a dose-response model to pool information across doses, such as in the CRM, is inevitably subject to the influence of model misspecification. As a result, on average, borrowing information across doses may not incur much performance gain. 

\newpage
\begin{center}
{\large  APPENDIX}
\end{center}
\begin{center}
{\em A. Derivation of $\lambda_{1j}$ and $\lambda_{2j}$ for the local optimal interval design}
\end{center}
We rewrite the decision error $\alpha(\lambda_{1j}, \lambda_{2j})$ given in equation (\ref{lerror}) as
$$
\alpha(\lambda_{1j}, \lambda_{2j})=\alpha_1(\lambda_{1j})+\alpha_2(\lambda_{2j})+\pi_{0j}+\pi_{1j},
$$
where 
\begin{eqnarray*}
\alpha_1(\lambda_{1j}) &=& \pi_{0j}Bin(n_j\lambda_{1j}; n, \phi) -\pi_{1j}Bin(n_j\lambda_{1j}; n, \phi _1)\\
\alpha_2(\lambda_{2j}) &=& \pi_{2j}Bin(n_j\lambda_{2j}-1; n, \phi_2)- \pi_{0j}Bin(n_j\lambda_{2j}-1; n, \phi).
\end{eqnarray*}
Therefore, in order to minimize $\alpha(\lambda_{1j}, \lambda_{2j})$, we can minimize $\alpha_1(\lambda_{1j})$ and $\alpha_2(\lambda_{2j})$ separately with regard to $\lambda_{1j}$ and $\lambda_{2j}$, respectively.
To minimize $\alpha_1(\lambda_{1j})$, let $b_{1j}=floor(n_j\lambda_{1j})$ and note that 
\begin{eqnarray*}
\alpha_1(\lambda_{1j})&=&\pi_{0j}Bin(n_j\lambda_{1j}; n_j, \phi)- \pi_{1j}Bin(n_j\lambda_{1j}; n_j, \phi _1)\\
 &= &\sum_{y=0}^{b_{1j}}{n_j \choose y}\{\pi_{0j}\phi^y(1-\phi)^{n_j-y}-\pi_{1j}\phi _1^y(1-\phi _1)^{n_j-y}\} \\
&=&\sum_{y=0}^{b_{1j}}\pi_{1j}{n_j \choose y}\phi _1^y(1-\phi _1)^{n_j-y}\left\{ \frac{\pi_{0j}}{\pi_{1j}}\left(\frac{\phi}{\phi _1}\right)^{y}\left( \frac{1-\phi}{1-\phi _1}\right)^{n_j-y}-1\right\}.
\end{eqnarray*}
Because $\phi>\phi _1$, $\left(\frac{\phi}{\phi _1}\right)^{y}\left( \frac{1-\phi}{1-\phi _1}\right)^{n_j-y}$ monotonically increases with $y$. Therefore, given ${\pi_{1j}}{n \choose y}\phi _1^y(1-\phi _1)^{n_j-y}>0$,  $\alpha_1(\lambda_{1j})$ is minimized when 
\begin{equation}
n_j \lambda_{1j} =  {\rm max} \left\{y:  \, \frac{\pi_{0j}}{\pi_{1j}}\left(\frac{\phi}{\phi _1}\right)^{y}\left( \frac{1-\phi}{1-\phi _1}\right)^{n_j-y} \le 1 \right\}. \label{condilocal}
\end{equation}
This leads to the solution
\begin{equation*} 
\lambda_{1j}  = \frac{\displaystyle {\rm log}\left(\frac{1-\phi _1}{1-\phi}\right)+ n_j ^{-1}{\rm log}\left(\frac{\pi_{1j}}{\pi_{0j}}\right)  }{\displaystyle{\rm log}\left(\frac{\phi(1-\phi _1)}{\phi_1(1-\phi)}\right)}. \label{lambda1} 
\end{equation*}

It is worth noting that the condition $\frac{\pi_{0j}}{\pi_{1j}}\left(\frac{\phi}{\phi _1}\right)^{y}\left( \frac{1-\phi}{1-\phi _1}\right)^{n_j-y} \le 1$ in equation (\ref{condilocal}) is equivalent to
$$ \frac{{\rm pr}(H_{0j}|y)}{{\rm pr}(H_{1j}|y)} \le 1$$
since
$$ {\rm pr}(H_{0j}|y) = \frac{{\rm pr}(H_{0j}) f(y|H_{0j})}{f(y)} = \frac{ \pi_{0j} {n \choose y} \phi^y(1-\phi)^{n_j-y}}{f(y)}   $$
$$ {\rm pr}(H_{1j}|y) = \frac{{\rm pr}(H_{1j}) f(y|H_{0j})}{f(y)} = \frac{ \pi_{0j} {n \choose y} \phi _1^y(1-\phi _1)^{n_j-y}}{f(y)}. $$
Therefore, $\lambda_{1j}$ is the boundary at which the posterior probability of $H_{1j}$ becomes more likely than that of $H_{0j}$.

In the same vein, it can be shown that 
$$\lambda_{2j} = \frac{{\rm log}(\frac{1-\phi}{1-\phi_2}) + n_j^{-1}{\rm log}(\frac{\pi_{0j}}{\pi_{2j}}) }{{\rm log}(\frac{\phi_2(1-\phi)}{\phi(1-\phi_2)})},$$
and $\lambda_{2j}$ is the boundary at which the posterior probability of $H_{2j}$ becomes more likely than that of $H_{0j}$.

\bigskip
\begin{center}
{\em B. Proof of Theorem 1}
\end {center}
 When $\pi_{0j}=\pi_{1j}=\pi_{2j}$, $\lambda_{1j}$ and $\lambda_{2j}$ are the likelihood-ratio hypothesis-testing boundaries, so $\lambda_{1j}<\phi<\lambda_{2j}$. Therefore, 
 \begin{eqnarray*}
  {\rm pr(dose \,\, escalation}|\hat{p}_j>\phi) &=& {\rm pr}(\hat{p}_j<\lambda_{1j}|\hat{p}_j>\phi)=0 \\
   {\rm pr(dose \,\, deescalation}|\hat{p}_j<\phi) &=& {\rm pr}(\hat{p}_j>\lambda_{2j}|\hat{p}_j<\phi)=0. 
\end{eqnarray*}

\newpage
\begin{center}
{\em C. Proof of Theorem 3}
\end {center}
{\small
\begin{eqnarray*}
&&{\rm \alpha_g(\lambda_{1j},\lambda_{2j})} \notag \\
 &=& {\rm pr}(H_{0j}) \int f(p_j|H_{0j}) {\rm pr}(\bar{\cal R}| p_j, H_{0j}) \,{\rm d} p_j+{\rm pr}(H_{1j}) \int f(p_j|H_{1j}) {\rm pr}(\bar{\cal E} | p_j, H_{1j})\,{\rm d} p_j \\
 &&+{\rm pr}(H_{2j}) \int f(p_j|H_{2j})  {\rm pr}(\bar{\cal D} |p_j,  H_{2j})\,{\rm d} p_j  \\
 &=& {\rm pr}(H _{0j}) \int f(p_j|H_{0j}) {\rm pr}(m_j \le n_j\lambda_{1j} \, {\rm or} \, m_j\ge n_j\lambda_{2j}  | p_j, H _{0j}) \,{\rm d} p_j \\
 &&+ {\rm pr}(H _{1j}) \int f(p_j|H_{0j}) {\rm pr}(m_j > n_j\lambda_{1j} | p_j, H _{1j}) \,{\rm d} p_j \\
 && +{\rm pr}(H _{2j}) \int f(p_j|H_2){\rm pr}(m_j < n_j\lambda_{2j} | p_j, H _{2j})\,{\rm d} p_j \notag  \\
 &=& {\rm pr}(H _{0j}) \sum_{y=0}^{b_{1j}} \int f(p_j|H_{0j})f(y|p_j, H_{0j}) \,{\rm d} p_j
 +{\rm pr}(H _{0j})-{\rm pr}(H _{0j})  \sum_{y=0}^{b_{2j}-1} \int f(p_j|H_{0j})f(y|p_j, H_{0j}) \,{\rm d} p_j \\
 && + {\rm pr}(H _{1j})-{\rm pr}(H _{1j})  \sum_{y=0}^{b_{1j}} \int f(p_j|H_{1j})f(y|p_j, H_{1j}) \,{\rm d} p_j 
 +{\rm pr}(H _{2j})  \sum_{y=0}^{b_{2j}-1} \int f(p_j|H_{2j})f(y|p_j, H_{2j}) \,{\rm d} p_j \\
 &=& {\rm pr}(H _{0j})+{\rm pr}(H _{1j}) + \sum_{y=0}^{b_{1j}} \left\{ {\rm pr}(H _{0j}) \int f(p_j|H_{0j})f(y|p_j, H_{0j}) \,{\rm d} p_j - {\rm pr}(H _{1j}) \int f(p_j|H_{1j})f(y|p_j, H_{1j}) \,{\rm d} p_j\right\}\\
 && + \sum_{y=0}^{b_{2j}-1} \left\{ {\rm pr}(H _{2j}) \int f(p_j|H_{2j})f(y|p_j, H_{2j}) \,{\rm d} p_j - {\rm pr}(H _{0j}) \int f(p_j|H_{0j})f(y|p_j, H_{0j}) \,{\rm d} p_j\right\}\\
 &=& {\rm pr}(H _{0j})+{\rm pr}(H _{1j}) + \sum_{y=0}^{b_{1j}} f(y) \{ {\rm pr}(H_{0j}|y)-{\rm pr}(H_{1j}|y)\} +
 \sum_{y=0}^{b_{2j}-1} f(y) \{{\rm pr}(H_{2j}|y)-{\rm pr}(H_{0j}|y)\}
\end{eqnarray*}
}
The last equality holds because
\begin{equation*}
{\rm pr}(H_{kj}|y) = \frac{{\rm pr}(H_{kj}) \int f(p_j|H_{kj}) f(y|p_j, H_{kj}) d p_j }{f(y)} , \quad k=0, 1, 2.
\end{equation*}
Therefore, in order to minimize $\alpha_g(\lambda_{1j},\lambda_{2j})$, we can minimize $\alpha_{g1}(\lambda_{1j})=\sum_{y=0}^{b_{1j}} f(y) \{ {\rm pr}(H_{0j}|y)-{\rm pr}(H_{1j}|y)\}$ and $\alpha_{g2}(\lambda_{2j})=\sum_{y=0}^{b_{2j}-1} f(y) \{{\rm pr}(H_{2j}|y)-{\rm pr}(H_{0j}|y)\}$ separately with regard to $\lambda_{1j}$ and $\lambda_{2j}$, respectively.

Now, 
\begin{eqnarray*}
{\rm pr}(H_{0j}|y) &=& \frac{{\rm pr}(H_{0j}) \int f(p_j|H_{0j}) f(y|p_j, H_{0j}) d p_j }{f(y)} \\
& =& \frac{ \pi_{0j} \int_{\phi_1}^{\phi_2} (\phi_2-\phi_1)^{-1} {n \choose y} p_j^y(1-p_j)^{n_j-y}\,{\rm d} p_j  }{f(y)} \\
&=&  \frac{\pi_{0j}\{Beta(\phi_2;j+1, n-j+1)-Beta(\phi _1;j+1, n-j+1)\}}{(\phi_2-\phi _1)(n_j+1)f(y)} 
\end{eqnarray*}
and similarly, it can be shown that
$${\rm pr}(H_{1j}|y) =\frac{\pi_{1j}Beta(\phi _1;j+1, n-j+1)}{\phi _1(n_j+1) f(y)}.$$

Therefore, 
\begin{eqnarray*}
\alpha_{g1}(\lambda_{1j})&=&\sum_{y=0}^{b_{1j}} \{ {\rm pr}(H_{0j}|y) -{\rm pr}(H_{1j}|y) \}  \\
 &= &\sum_{y=0}^{b_{1j}} {\rm pr}(H_{1j}|y) \left\{ \frac{{\rm pr}(H_{0j}|y)}{{\rm pr}(H_{1j}|y)}-1 \right\}  \\
 &=&\sum_{y=0}^{b_{1j}} {\rm pr}(H_{1j}|y) \left\{ \frac{\pi_{0j}\phi _1}{\pi_{1j}(\phi_2-\phi _1)} \frac{Beta(\phi_2;y+1, n_j-y+1)}{Beta(\phi _1;y+1, n_j-y+1)}
-\frac{1}{\pi_{1j}(\phi_2-\phi _1)} -1 \right\}.
\end{eqnarray*}
 Because ${\rm pr}(H_{1j}|y) \ge 0$, in order to establish that $\alpha_1(\lambda_{1j})$ is minimized when
\begin{equation}
n_j \lambda_{1j} =  {\rm max} \left\{y:  \, \frac{{\rm pr}(H_{0j}|y)}{{\rm pr}(H_{1j}|y)} \le 1 \right\}, \label{pfmin}
\end{equation}
we need to show only that $\frac{\displaystyle Beta(\phi_2; y+1, n-y+1)}{\displaystyle Beta(\phi _1; y+1, n-y+1)}$ is a monotonic increasing function of $y$. Letting $y^*=y+1$, and using the following property of the cumulative density function (CDF) of the Beta distribution, 
$$
Beta(x, a, b) = \sum_{j=a}^{a+b-1} \frac{(a+b-1)!}{j!(a+b-1-j)!} x^j(1-x)^{a+b-1-j}
$$
when $a$ and $b$ are integers, we have 
$$
Beta(\phi_2; y+1, n_j-y+1) = Beta(\phi_2; y^*+1, n_j-y^*+1) + \frac{(n+1)!}{y!(n+1-y)!} \phi_2^y (1-\phi_2)^{n_j+1-y}
$$
where, as in our case, $y$ and $n_j$ are integers. Let  $C(y) = {(n+1)!}/\{y!(n+1-y)!\}$, then  it follows that
\begin{eqnarray*}
\frac{Beta(\phi_2;y+1, n_j-y+1)}{Beta(\phi _1;y+1, n_j-y+1)} &=& \frac{Beta(\phi_2;y^*+1, n_j-y^*+1) + C(y) \phi_2^y (1-\phi_2)^{n_j+1-y}}{Beta(\phi _1; y^*+1, n_j-y^*+1) + C(y) \phi _1^y (1-\phi _1)^{n_j+1-y}} \\
&=& \frac{Beta(\phi_2;y^*+1, n_j-y^*+1)}{Beta(\phi _1;y^*+1, n_j-y^*+1)}  \frac{1 + \frac{\displaystyle C(y) \phi_2^y (1-\phi_2)^{n_j+1-y}}{\displaystyle Beta(\phi_2; y^*+1, n_j-y^*+1)}}{1 + \frac{\displaystyle C(y) \phi _1^y (1-\phi _1)^{n_j+1-y}}{\displaystyle Beta(\phi _1; y^*+1, n_j-y^*+1)}}\\
&=& \frac{Beta(\phi_2;y^*+1, n_j-y^*+1)}{Beta(\phi _1;y^*+1, n_j-y^*+1)} \times \frac{\displaystyle 1 + \frac{C(y) \phi_2^y (1-\phi_2)^{n_j+1-y}}{\displaystyle \sum\limits_{k=y^*+1}^{n_j+1} C(k) \phi_2^k (1-\phi_2)^{n_j+1-k} }}{1 + \frac{\displaystyle C(y) \phi _1^y (1-\phi _1)^{n_j+1-y}}{\displaystyle 
\sum\limits_{k=y^*+1}^{n_j+1} C(k) \phi _1^k (1-\phi _1)^{n_j+1-k} }}\\
&=& \frac{Beta(\phi_2;y^*+1, n_j-y^*+1)}{Beta(\phi _1;y^*+1, n_j-y^*+1)} \times \frac{1+ (\sum_{k=y^*+1}^{n_j+1} \frac{C(k)}{C(y)}(\frac{\phi_2}{1-\phi_2})^{k-y})^{-1}}{1+ (\sum_{k=y^*+1}^{n_j+1} \frac{C(k)}{C(y)}(\frac{\phi _1}{1-\phi _1})^{k-y})^{-1}}.\\
\end{eqnarray*}
Now because $\phi_2>\phi _1$, it follows that
$$
\frac{Beta(\phi_2;y+1, n_j-y+1)}{Beta(\phi _1;y+1, n_j-y+1)}<\frac{Beta(\phi_2;y^*+1, n_j-y^*+1)}{Beta(\phi _1;y^*+1, n_j-y^*+1)}.
$$
Therefore, $\frac{\displaystyle Beta(\phi_2; y+1, n-y+1)}{\displaystyle Beta(\phi _1; y+1, n-y+1)}$ is a monotonic increasing function of $y$, and equation (\ref{pfmin}) follows.

Along the same line, we can  prove that $\alpha_{g2}(\lambda_{2j})$ is minimized when
\begin{equation*}
n_j \lambda_{2j} =  {\rm max} \left\{y:  \, \frac{{\rm pr}(H_{2j}|y)}{{\rm pr}(H_{0j}|y)} \le 1 \right\}. \label{condi}
\end{equation*}

 \bigskip
\bigskip
\begin{center}
{\em D. Proof of Theorem 5}
\end {center}
 It is known that Bayesian hypothesis testing is consistent in the sense that if $H_{kj}$ is true, ${\rm pr}(H_{kj}|n_j)\to 1$ when $n_j \to \infty$ (Walker and Hjort, 2001). Now, according to Theorem 3, the values of $\lambda_{1j}$ and $\lambda_{2j}$ that minimize the global decision error rate (\ref{gerror}) are  the boundaries at which the posterior probabilities of $H_1$ and $H_2$, respectively, become more likely than that of $H_0$.Therefore, when $n_j\to \infty$,  $\lambda_{1j} \to \phi_1$ and $\lambda_{2j} \to \phi_2$ as both $\lambda_{1j}$ and $\lambda_{2j}$ converge to constants. The proof provided by Oron, Azriel and Hoff (2011) can be directly used to obtain the result.

\vspace{1in}
\begin{center}
\large{REFERENCES}
\end{center}

\begin{description}
\item Ahn, C.  (1998) An evaluation of phase I cancer clinical trial designs. {\em Statistics in Medicine}, 17, 1537-1549.

\item Azriel, D. (2012). A note on the robustness of the continual reassessment method. {\it Statistics \& Probability
Letters},  { 82}, 902-906.

\item Babb, J., Rogatko, A., and Zacks, S. (1998) Cancer phase I clinical trials: efficient dose
escalation with overdose control,  {\em Statistics in Medicine}, { 17}, 1103-1120.

\item  Barlow, R.E., Bartholomew, D.J., Bremner, J.M., and Brunk, H.D. (1972) {\em Statistical Inference
under Order Restrictions}, John Wiley \& Sons, London, New York.

\item Cheung, Y. (2005) Coherence principles in dose-finding studies. {\em Biometrika}, { 92}, 863-873.

\item Chevret, S. (2006) {\em Statistical Methods for Dose-Finding
Experiments}, John Wiley \& Sons, London.

\item Durham, S.D. and Flournoy, N. (1994) Random walks for quantile estimation. In: Berger, J.O., Gupta, S.S. (Eds.), {\em Statistical Decision Theory and Related Topics} Springer, New York, pp. 467Ð476.

\item Gerke, O. and Siedentop, H. (2008) Optimal phase I dose-escalation trial designs in oncology---a simulation study.
{\em Statistics in Medicine}, { 27}, 5329-5344.

\item  Gezmu M. and Flournoy, N. (2006) Group up-and-down designs for dose-finding. {\it Journal of Statistical Planning and Inference}, 136, 1749-1764.

\item
Iasonos, A., Wilton, A.S., Riedel, E.R., Seshan, V.E. and Spriggs, D.R. (2008) A comprehensive comparison of the continual reassessment method to the standard 3 + 3 dose escalation scheme in phase I dose-finding studies. {\em Clinical Trials}, { 5}, 465-477.

\item
Ivanova, A., Flournoy, N., and Chung, Y. (2007) Cumulative cohort design for dosefinding.
{\em Journal of Statistical Planning and Inference}, 137, 2316-2317.

\item
Ivanova, A. and Kim, S. (2009) Dose-finding for binary ordinal and continuous outcomes with monotone objective function: a unified approach. {\em Biometrics}, { 65}, 307-315.

\item
Ji, Y., Liu, P., Li, Y., and Bekele, B. N. (2010) A modified toxicity probability interval method for dose-finding trials. {\em Clinical Trials}, { 7}, 653-663.

\item  Korn, E.L., Midthune, D., Chen, T.T., Rubinstein, L.V., Christian, M.C., and Simon, R.M.(1994)  A comparison of two phase I trial
designs. {\it Statistics in Medicine}, { 13}, 1799-1806.

\item  Lee, S. and Cheung, Y. (2009). Model calibration in the continual reassessment method.
{\it Clinical Trials}; { 6}, 227-238.

\item Lee, S. and Cheung, Y. (2011). Calibration of prior variance in the Bayesian continual
reassessment method. {\it Statistics in Medicine}; { 30}, 2081-2089.

\item Leung, D. and Wang, Y.G. (2001) Isotonic designs for phase I trials. {\em Controlled Clinical Trials}, { 22}, 126-138.

\item  Lin, Y. and Shih, W.J. (2001) Statistical properties of the traditional
algorithm-based designs for phase I cancer clinical trials.
{\em Biostatistics},  { 2},  203-215.

\item Oron, A., Azriel, D. and Hoff, P. (2011) Dose-finding designs: the role of convergence properties.
{\em The International Journal of Biostatistics},  { 7}, Article 39.

\item O'Quigley, J., Pepe, M., and Fisher, L. (1990) Continual
reassessment method: a practical design for phase 1 clinical trials
in cancer, {\em Biometrics}, { 46}, 33-48.

\item Paoletti, X. , O'Quigley, J. and Maccario, J. (2004). Design efficiency in dose finding studies. 
{\it Computational Statistics \& Data Analysis}; { 45}, 197-214.

\item Storer, B. E.  (1989). Design and Analysis of Phase I
Clinical Trials. {\em Biometrics}, { 45}, 925-937.

\item Ting, N. (2006) {\em Dose Finding in Drug Development}, Springer: Cambridge.

\item Tsutakawa, R. (1967)  Random walk design in bio-assay. {\em Journal of American Statistical Association}, { 62}, 842-856.

\item Walker, S.G. and Hjort, N.L. (2001). On Bayesian consistency, {\it Journal of the Royal Statistical Society, Series B}, { 63}, 811-821.

\item Wetherill, G. (1963) Sequential estimation of quantal response curves, {\em Journal of Royal Statistical Society Series B}, { 25}, 1-48.

\item Whitehead, J. and Brunier, H. (1995) Bayesian decision procedures for dose determining experiments, {\em Statistics in Medicine},14, 885-893.



\item Yuan, Z. and Chappell, R. (2004). Isotonic designs for phase I cancer clinical trials with multiple
risk groups, {\em Clinical Trials} { 1}, 499-508.

\end{description}

\newpage
\clearpage

\newpage
\clearpage

\newpage
\clearpage

\newpage
\begin{table}
\begin{center}
{\small
\caption{Values of $\lambda_{1j}$ and $\lambda_{2j}$ under the local optimal interval design for different target toxicity rates when $\phi_1=0.6\phi$, $\phi_2=1.4\phi$ and $\pi_{0j}=\pi_{1j}=\pi_{2j}$. \label{tablocal}}}
\begin{tabular}{cccccccccccc}\hline\hline
Interval & \multicolumn{6}{c}{Target toxicity rate $\phi$} \\
\cline{2-8} 
boundaries && 0.15 & 0.2 & 0.25 & 0.3 & 0.35 & 0.4 \\ 
\hline
$\lambda_{1j}$  && 0.118 & 0.157 & 0.197 & 0.236 & 0.276 & 0.316 \\
$\lambda_{2j}$  && 0.179 & 0.238 & 0.298 & 0.358 & 0.419 & 0.479 \\
\hline\hline
\end{tabular}
\end{center}
\end{table}

\begin{table}[H]
\begin{center}
{\small
\caption{Interval  boundaries and dose elimination boundaries for the global optimal designs when the target toxicity rate $\phi=0.25$ with $\phi_1=0.15$ and $\phi_2=0.35$. \label{tabglobal}}
\begin{tabular}{lcccccccccccccccc}\hline\hline
&&\multicolumn{12}{c}{Number of cumulative patients treated at the current dose ($n_j$)} \\
\cline{2-16} 
Boundary & 1 & 2 & 3 & 4 & 5 & 6 & 7 & 8  & 9 & 10 & 11 & 12 & 13 & 14 & 15 \\
\hline
 $\lambda_{1j}$ & 0/1 &  0/2 &  0/3 & 0/4 & 0/5 & 0/6 & 0/7 & 1/8 & 1/9 &  1/10 &  1/11 &  1/12 &  1/13 &  1/14 &  2/15 \\
$\lambda_{2j}$ & 1/1 &   2/2 & 2/3 & 2/4 & 3/5 & 3/6 & 4/7 & 4/8 & 5/9 &  5/10 &  5/11 &  6/12 &  6/13 &  7/14 &  7/15  \\
Elimination & NA & NA & 3/3 & 3/4 & 3/5 & 4/6 & 4/7 & 4/8 & 5/9 &  5/10 &  6/11 &  6/12 &  6/13 &  7/14 &  7/15 \\
\hline\hline
\end{tabular}
}
\end{center}
\end{table}

\clearpage

\newpage
\begin{table}
\begin{center}
{ \small
\caption{Simulation results when the dose-toxicity scenarios are randomly generated. \label{OC}}
\begin{tabular}{lcccccccccccc}\hline\hline
 & Selection $\%$ & $ \%$ of patients  &Average & Risk of & Risk of &  Sample  \\
Design & of MTD &  at MTD & toxicity rate  & poor allocation & high toxicity & size  \\ \hline
\multicolumn{7}{c}{Average probability difference around the target = 0.1} \vspace{1mm} \\
GUD & 45.2 &		{\bf 29.8} &18.6&	{\bf 21.1} &		{\bf 14.5} & 	35.3 \\
CCD & 46.0 &		35.2 &19.8 &	39.1 &		16.2 &	35.3 \\
mTPI & 44.8 &		35.3 &19.8 &	39.9 &		16.9 &	35.2 \\
CRM & 44.7 &		35.4 &19.3 &	43.0 &		16.9 &	35.2\\
Global optimal & 45.4 &		35.5 &19.9&	41.7 &	 	17.6 &	35.2\\
Local optimal & 46.2 &		33.0 &19.2 &	{\bf 28.7} &		{\bf 14.8} &	35.3\\
\\
\multicolumn{7}{c}{Average probability difference around the target = 0.07} \vspace{1mm} \\
GUD & 36.6 &		26.3 &19.3 &	{\bf 29.4} &		{\bf 15.9} &	35.2\\
CCD & 36.2 &		29.6 &20.0 &	49.4 &		16.5  &	35.1\\
mTPI & 36.1 &		29.8 &20.0 &	50.7 &		17.3 &	35.1 \\
CRM & 33.3 &		28.6 &19.5 &	54.7 &		17.5 &	35.1\\
Global optimal & 36.6 &		29.6 &20.1 &	51.5 &		18.0 &	35.1\\
Local optimal &37.0 &		28.5 &19.5 &	{\bf 38.6} &		{\bf 15.5} &	35.1\\
\\
\multicolumn{7}{c}{Average probability difference around the target = 0.15}\vspace{1mm}  \\
GUD & 56.8 &		{\bf 34.8} &18.3 &	{\bf 10.9} &		{\bf 13.6} &	35.4\\
CCD & 57.7 &		43.4 &19.8 &	26.0 &		16.8 &	35.4\\
mTPI & 57.6 &		44.0 &20.1 &	26.9 &		18.1 &	35.4 \\
CRM & 59.8 &		44.7 &19.6 &	26.6 &		17.9 &	35.4\\
Global optimal & 57.5 &		44.1 &20.1 &	27.7 &		18.1 &	35.4 \\
Local optimal &57.6 &		43.4 &19.0  &	{\bf 18.0} &		{\bf 15.5} &	35.3\\
\hline\hline
\end{tabular}
}
\end{center}
\end{table}

\newpage
\begin{table}
\begin{center}
{ \small
\caption{Simulation results under six prespecified dose-toxicity scenarios. The target toxicity rate of 0.25 is underlined. \label{fixedsimu}}
\begin{tabular}{lcccccccccccc}\hline\hline
&&\multicolumn{6}{c}{Dose level} & Risk of & Risk of \\
\cline{3-8} Design && 1 & 2 & 3 & 4 & 5 & 6 &  poor allocation & high toxicity  \\ \hline

Scenario 1  &Pr(toxicity)& \underline{0.25} & 0.35 & 0.5 & 0.6 & 0.7 & 0.8  & &\\

GUD & Selection (\%) & 57.3 & 25.0 & 3.4 & 0.3 & 0.0 & 0.0 &{ 5.2} & 60.7  \\ 
& \# patients &  21.0 & 9.2 & 2.4 & 0.3 & 0.0 & 0.0 & &  \\

CCD & Selection (\%) & 60.8 & 23.0 & 1.5 & 0.0 & 0.0 & 0.0  & 19.8 & 52.8  \\
& \# patients & 22.4 & 8.8 & 1.4 & 0.1 & 0.0 & 0.0 & &  \\


mTPI & Selection  (\%) & 58.2 & 25.3 & 1.8 & 0.1 & 0.0 & 0.0 & 23.6 & 53.7  \\
& \# patients & 21.8 & 9.3 & 1.5 & 0.1 & 0.0 & 0.0  & & \\

CRM  & Selection  (\%) & 55.6 & 26.2 & 1.1 & 0.0 & 0.0 & 0.0 & 22.9 & 52.9  \\
&\# patients &21.9 & 9.3 & 1.4 & 0.1 & 0.0 & 0.0 & &  \\

Global optimal & Selection  (\%) & 59.4 & 24.5 & 1.6 & 0.0 & 0.0 & 0.0 & 24.8 & 54.1  \\
&\# patients &21.5 & 9.5 & 1.5 & 0.1 & 0.0 & 0.0  & &   \\

Local optimal & Selection  (\%) & 63.0 & 20.6 & 1.6 & 0.1 & 0.0 & 0.0 & { 13.8} & 53.4  \\
&\# patients &22.9 & 8.0 & 1.7 & 0.2 & 0.0 & 0.0 & &   \\
\\

Scenario 2  &Pr(toxicity)& 0.03 &  0.06 & 0.1 & \underline{0.25} & 0.35 & 0.5  & &\\

GUD & Selection (\%) & 0.0 & 1.1 & 21.6 & 52.4 & 21.0 & 3.9 & { 27.4} & { 0.5} \\ 
& \# patients &  4.4 & 6.7 & 10.1 & 9.6 & 4.1 & 1.1 & &  \\

CCD & Selection (\%) & 0.0 & 1.1 & 19.1 & 54.3 & 22.6 & 2.8  & 35.4 & 5.0  \\
& \# patients & 3.7 & 4.9 & 8.8 & 12.4 & 5.2 & 1.0  & &  \\


mTPI & Selection  (\%) & 0.0 & 1.1 & 18.1 & 53.4 & 24.1 & 3.3  & 37.3 & 6.4  \\
& \# patients & 3.7 & 4.8 & 8.5 & 12.5 & 5.4 & 1.1  & & \\

CRM  & Selection  (\%) & 0.0 & 2.8 & 26.1 & 56.4 & 14.1 & 0.6  & 40.4 & 4.4  \\
&\# patients &3.7 & 5.2 & 10.7 & 12.0 & 3.8 & 0.5 & &  \\

Global optimal & Selection  (\%) & 0.0 & 1.3 & 18.8 & 52.3 & 24.5 & 3.1 & 38.6 & 7.2  \\
&\# patients &3.7 & 4.8 & 8.3 & 12.5 & 5.6 & 1.1  & &   \\

Local optimal & Selection  (\%) & 0.0 & 1.0 & 21.3 & 55.1 & 20.5 & 2.1  & { 17.7} & { 3.2}  \\
&\# patients &4.0 & 5.3 & 9.3 & 11.5 & 4.7 & 1.2 & &   \\
\\

Scenario 3  &Pr(toxicity)&   0.01 & 0.04 & 0.06 & 0.1 & \underline{0.25} & 0.35   & &  \\

GUD & Selection (\%) & 0.0 & 0.0 & 2.0 & 22.0 & 54.0 & 22.0 & { 35.0} & 0.0 \\ 
& \# patients &  3.6 & 4.7 & 6.1 & 9.6 & 8.2 & 3.9  & &  \\

CCD & Selection (\%) & 0.0 & 0.0 & 2.0 & 19.0 & 58.0 & 21.0 & 40.0 & 1.0  \\
& \# patients & 3.1 & 4.0 & 5.2 & 8.0 & 11.0 & 4.8 & &  \\


mTPI & Selection  (\%) & 0.0 & 0.0 & 2.0 & 19.0 & 57.0 & 22.0 & 41.0 & 2.0  \\
& \# patients & 3.1 & 4.0 & 5.1 & 7.7 & 11.0 & 5.0 & & \\

CRM  & Selection  (\%) & 0.0 & 0.0 & 5.0 & 34.0 & 52.0 & 9.0 & 51.0 & 0.0  \\
&\# patients &3.1 & 3.7 & 7.2 & 9.4 & 9.7 & 2.9 & &  \\

Global optimal & Selection  (\%) & 0.0 & 0.0 & 3.0 & 18.0 & 56.0 & 23.0  & 41.0 & 2.0  \\
&\# patients &3.1 & 4.0 & 5.2 & 7.6 & 11.1 & 5.0 & &   \\

Local optimal & Selection  (\%) & 0.0 & 0.0 & 3.0 & 21.0 & 65.0 & 11.0  & { 31.0} & 1.0  \\
&\# patients &3.5 & 4.3 & 5.2 & 9.0 & 10.2 & 3.7 & &   \\
\hline\hline
\end{tabular}
}
\end{center}
\end{table}

\begin{table}
\begin{center}
{\small
Table 4 continued. 
\begin{tabular}{lcccccccccccc}\hline\hline
&&\multicolumn{6}{c}{Dose level} & Risk of & Risk of \\
\cline{3-8} Design && 1 & 2 & 3 & 4 & 5 & 6 &  poor allocation & high toxicity \\ \hline
%
%
%
%
%
%
%

Scenario 4  &Pr(toxicity)& 0.05 & 0.1 & \underline{0.25} & 0.32 & 0.5 & 0.6  & &  \\

GUD & Selection (\%) & 0.4 & 18.9 & 49.2 & 26.9 & 4.2 & 0.4 &  { 19.1} & { 2.4} \\ 
& \# patients &  6.8 & 11.9 & 10.4 & 5.3 & 1.4 & 0.2 & &  \\

CCD & Selection (\%) & 0.5 & 18.8 & 50.4 & 27.4 & 2.8 & 0.2 &  34.0  & 13.2  \\
& \# patients & 4.5 & 9.5 & 14.0 & 6.5 & 1.4 & 0.1  & &  \\


mTPI & Selection  (\%) & 0.5 & 18.0 & 49.1 & 29.0 & 3.1 & 0.2 &  36.3  & 15.6  \\
& \# patients & 4.5 & 9.1 & 14.1 & 6.7 & 1.4 & 0.1  & & \\

CRM  & Selection  (\%) & 0.1 & 18.1 & 61.1 & 19.5 & 1.1 & 0.0 &  31.3  & 14.2  \\
&\# patients &4.4 & 9.7 & 15.7 & 5.2 & 0.8 & 0.1 & &  \\

Global optimal & Selection  (\%) & 0.7 & 18.3 & 49.0 & 28.8 & 3.0 & 0.2 &  37.7 & 16.8  \\
&\# patients &4.5 & 9.0 & 14.1 & 6.9 & 1.5 & 0.1 & &   \\

Local optimal & Selection  (\%) & 0.4 & 19.0 & 53.0 & 24.7 & 2.8 & 0.1 &  { 27.8} & { 9.8}  \\
&\# patients &5.1 & 10.2 & 13.2 & 5.9 & 1.6 & 0.2  & &   \\
\\

Scenario 5  &Pr(toxicity)& 0.01 & 0.02 & 0.03 & 0.04 & 0.05 & \underline{0.25} & &  \\

GUD & Selection (\%) & 0.0 & 0.0 & 0.1 & 0.5 & 17.0 & 82.5 & 19.4 & 0.0 \\ 
& \# patients &  3.3 & 3.5 & 3.8 & 4.7 & 9.6 & 11.0 & &  \\

CCD & Selection (\%) & 0.0 & 0.0 & 0.1 & 0.5 & 15.2 & 84.2  & 11.6 & 0.0  \\
& \# patients & 3.2 & 3.4 & 3.7 & 3.9 & 6.2 & 15.6 & &  \\


mTPI & Selection  (\%) & 0.0 & 0.0 & 0.1 & 0.5 & 14.3 & 85.0 & 11.6 & 0.0  \\
& \# patients & 3.2 & 3.4 & 3.7 & 3.9 & 6.0 & 15.8 & & \\

CRM  & Selection  (\%) & 0.0 & 0.1 & 2.1 & 9.8 & 18.1 & 69.9 & 33.9 & 0.0  \\
&\# patients &3.2 & 3.5 & 4.7 & 5.1 & 7.0 & 12.5 & &  \\

Global optimal & Selection  (\%) & 0.0 & 0.0 & 0.2 & 0.5 & 14.5 & 84.8  & 11.6 & 0.0  \\
&\# patients &3.2 & 3.4 & 3.7 & 3.9 & 5.8 & 16.0 & &   \\

Local optimal & Selection  (\%) & 0.0 & 0.0 & 0.1 & 0.7 & 16.8 & 82.4 & 14.1 & 0.0  \\
&\# patients &3.3 & 3.5 & 3.8 & 4.0 & 7.6 & 13.8 & &   \\

\hline\hline
\end{tabular}
}
\end{center}
\end{table}
\end{document}